\documentclass[%
reprint,
superscriptaddress,
amsmath,amssymb,
aps,longbibliography,
]{revtex4-2}

\usepackage{graphicx}
\usepackage{subcaption}
\usepackage{verbatim}
\usepackage[version=4]{mhchem}
\usepackage{xcolor}
\usepackage{physics}
\usepackage[normalem]{ulem}
\usepackage{pdfpages}
\usepackage{pgffor}
\usepackage{float}
\usepackage{booktabs}
\usepackage{tikz}
\usepackage{cancel}
\makeatletter
\AtBeginDocument{\let\LS@rot\@undefined}
\makeatother

\begin{document}

\title{Overcoming the limitations of body-ordered potentials for atomistic machine learning}

\author{Filippo Bigi}
\email{filippo.bigi@epfl.ch}
\affiliation{Laboratory of Computational Science and Modeling, Institut des Mat\'eriaux, \'Ecole Polytechnique F\'ed\'erale de Lausanne, 1015 Lausanne, Switzerland}

\author{Michelangelo Domina}
\affiliation{Laboratory of Computational Science and Modeling, Institut des Mat\'eriaux, \'Ecole Polytechnique F\'ed\'erale de Lausanne, 1015 Lausanne, Switzerland}

\author{Maximilian Lloyd Ach}
\affiliation{Laboratory of Computational Science and Modeling, Institut des Mat\'eriaux, \'Ecole Polytechnique F\'ed\'erale de Lausanne, 1015 Lausanne, Switzerland}
\affiliation{Theory Department, Fritz Haber Institute of the Max Planck Society, 14195 Berlin, Germany}

\author{Sergey N. Pozdnyakov}
\affiliation{Laboratory of Computational Science and Modeling, Institut des Mat\'eriaux, \'Ecole Polytechnique F\'ed\'erale de Lausanne, 1015 Lausanne, Switzerland}

\author{Paolo Pegolo}
\affiliation{Laboratory of Computational Science and Modeling, Institut des Mat\'eriaux, \'Ecole Polytechnique F\'ed\'erale de Lausanne, 1015 Lausanne, Switzerland}

\author{Joseph W. Abbott}
\affiliation{Laboratory of Computational Science and Modeling, Institut des Mat\'eriaux, \'Ecole Polytechnique F\'ed\'erale de Lausanne, 1015 Lausanne, Switzerland}

\author{Kevin K. Huguenin-Dumittan}
\affiliation{Laboratory of Computational Science and Modeling, Institut des Mat\'eriaux, \'Ecole Polytechnique F\'ed\'erale de Lausanne, 1015 Lausanne, Switzerland}

\author{Michele Ceriotti}
\affiliation{Laboratory of Computational Science and Modeling, Institut des Mat\'eriaux, \'Ecole Polytechnique F\'ed\'erale de Lausanne, 1015 Lausanne, Switzerland}

\newcommand{\FB}[1]{{\color{teal}{#1}}}
\newcommand{\MC}[1]{{\color{blue}#1}}
\newcommand{\chatgpt}[1]{{\color{orange}{#1}}}
\newcommand{\MD}[1]{{\color{purple}{#1}}}
\newcommand{\JA}[1]{{\color{green}{#1}}}

\date{\today}%

\begin{abstract}
Machine-learning interatomic potentials have become indispensable tools in atomistic simulations.
The most popular models rely heavily on physical priors, such as locality, smoothness and symmetry, that are assumed to improve the accuracy and transferability of the trained models. 
However, recent evidence suggests that, in the data-rich regime, unconstrained models that learn symmetry from the data can achieve very competitive accuracy and computational efficiency, without sacrificing stability and generalization power.
Many equivariant symmetric architectures historically rely on the assumption that the interatomic potential can be approximated well by a convergent \emph{cluster expansion}, i.e., a decomposition in terms associated to atomic pairs, triplets, quadruplets, and higher-order tuples.
We describe an architecture that incorporates locality, smoothness, and symmetry priors, but that -- thanks to a computationally efficient tensor product formulation -- does not assume low-order truncation of the body-ordered series.

\end{abstract}

\maketitle

\section{Introduction}

In recent years, the use of machine learning (ML) in the physical sciences has greatly enhanced the modeling of atomic-scale phenomena and contributed to advancing their understanding~\cite{butl+18nature}. In particular, ML models have been used to approximate quantum mechanical observables at a fraction of the computational cost required for \textit{ab initio} calculations, making it possible to predict the microscopic properties of molecules and materials in a cheap and scalable way~\cite{unke2021machine,huang2021ab}.

Applications of ML in this domain have traditionally strived to incorporate prior knowledge about the physics being modelled (e.g. in terms of the exact mathematical representation of Euclidean symmetries~\cite{behl-parr07prl}), under the assumption that doing so leads to improved transferability and data efficiency~\cite{botnet,mace,grace}.
Considerations on the locality of quantum mechanical treatments of condensed-phase systems~\cite{prod-kohn05pnas} underpin the choice of local graph-based models as the most popular architecture, with long-range extensions being usually built around physically-motivated power-law asymptotics~\cite{hugu+23jpcl,grasselli2026long}. 
Another widespread assumption is that interactions can be approximated as a rapidly-converging \emph{body-ordered expansion}~\cite{thomas2022body}, i.e. as a series of terms depending exclusively on pairs, triplets, quadruplets of atoms.
This assumption is grounded in pseudopotential theory~\cite{Moriarty_1988}, motivates cluster expansion theory of alloys~\cite{sanc+84pa} and very successful molecular interatomic potentials~\cite{medd+14jctc}, and is implicitly or explicitly invoked by the \emph{atomic cluster expansion}~\cite{drau19prb} (ACE), as well as by many of the equivariant neural network inspired by it~\cite{nequip,mace,grace}.
However, fast convergence of the cluster expansion depends on the type of bonding~\cite{Hermann_2007}, and on the definition of an appropriate, environment-dependent atomic reference.
A ``vacuum reference'', which is appropriate when describing matter across many different density regimes, often leads to oscillatory, non-convergent behavior~\cite{tan2025convergence,chong2025resolving}.
This is especially problematic when pursuing ``universal'' models, which in principle aim to afford good off-the-shelf accuracy on any chemical composition and geometry. Indeed, several recent efforts have aimed to achieve higher transferability by training on large general-purpose datasets containing tens or hundreds of millions of diverse atomic structures and billions of labels~\cite{esen,sevennet-omni,mace-mh,grace-foundation,pushing-pet,uma,orb,dpa3,new-nequip}.

In this work we strive to overcome these limitations, by incorporating physical considerations in the definition of short-range discretizations of the descriptors, reflecting the exponential spatial decay of interactions based on atomic orbitals, but we let the body order grow to the typical number of atoms within the receptive field of the model, using an efficient equivariant tensor product that allows to reach this limit without a dramatic increase in computational effort.
We call this architecture Smooth Physical Architecture with Compact Equivariants, to allude to the fact that---despite the obvious similarities with the plethora of models inspired by ACE---it is not a cluster expansion, because it effectively incorporates \emph{all} orders of interactions.

\section{Theory}

\subsection{Equivariant atomistic machine learning}

When approximating quantum mechanical observables with machine-learning models, the microscopic properties of interest must be predicted from the positions and chemical species of the atoms in the corresponding structure. In this context, enforcing the correct underlying physical symmetries is desirable in the sense that 1) it guarantees that the trained model obeys them exactly and 2) it removes the need for the model to (approximately) learn the transformation rules of the target(s) under translations, rotations, inversion and atom exchanges.

While translational and atom-exchange symmetries are almost invariably enforced through the calculation of relative atomic positions and permutation-invariant pooling operations~\cite{duval2023hitchhiker}, symmetries under rotation and inversion are commonly imposed via O(3)-equivariant operations~\cite{nequip,mace,allegro,grace} (although a growing number of architectures are not enforcing these constraints~\cite{pet,escaip,orb}).

A common first operation in O(3)-equivariant models is therefore the expansion of relative atomic positions $\mathbf{r}_{ij}$ in a spherical basis:
\begin{equation}
    u_{nlm}(\mathbf{r}_{ij}) = R_{nl}(r_{ij}) Y_{lm}(\mathbf{\hat{r}}_{ij}),
\end{equation}
where the spherical harmonics $Y_{lm}(\mathbf{\hat{r}})$ naturally appear as basis functions of the irreducible representations of the O(3) group, and $R_{nl}(r)$ are radial basis functions.

\subsection{Short-range physical priors}

Aside from encoding the underlying fundamental physical symmetries, it is common practice to include other priors into the functional forms of atomistic machine learning models with the aim of increasing their accuracy and data-efficiency.
\begin{itemize}
\item Smoothness: generally, ground-state properties (which one often aims to approximate) are smooth. This is a consequence of the non-crossing rule~\cite{non-crossing}, which can only be broken in the case of symmetric configurations. In practice, this translates to the use of functional forms that can be differentiated infinitely many times. Smoothness is also very closely related to regularization, with several approaches attempting to regularize the corresponding models according to some mathematical definition of smoothness~\cite{darby2026regularity}.
\item Locality and message passing: in a sense, this is the assumption that closer atoms will influence one another more strongly. Often justified with reference to the ``short-sightedness'' of interatomic interactions~\cite{prod-kohn05pnas}, this physical prior is usually implemented as a finite interaction cutoff radius (possibly combined with some form of message-passing scheme in graph neural networks).
\item Body-ordering: the body-order expansion of quantum-mechanical properties was often believed to converge until recently. Indeed, even though the body-ordered expansion usually converges quickly at the molecular level, or for datasets that are sufficiently homogeneous to define an effective atomic limit of the expansion, the most recent evidence indicates that the vacuum expansion either converges slowly~\cite{thomas2022body} (possibly preventing existing machine-learning models from correctly capturing body order) or diverges~\cite{tan2025convergence,chong2025resolving}. Nevertheless, it has been observed empirically that body-ordered models tend to generalize better to unfamiliar structures~\cite{botnet,grace}.
\end{itemize}

These priors have been used successfully in a wide range of atomistic models. However, the fundamental nature of short-range interatomic interactions, and especially its behavior with distance, is often neglected in the design of such architectures in favor of other heuristics~\cite{darby2026regularity}. In what follows, we investigate how a model can incorporate some of the mathematical properties of short-range interactions.

\subsection{Smoothness and regularization}\label{sec:smoothness-and-regularization}

A number of works~\cite{jinnouchi2019fly,kocer2019novel,bigi2022smooth} have used a basis of spherical Bessel functions as the radial basis to be employed in atomistic machine learning. These basis functions correspond to the radial part of the functions $u(\boldsymbol{r})$ that minimize the Rayleigh quotient
\begin{equation}\label{eq:rayleigh}
    R = \frac{\int_{\Omega} |\nabla u(\boldsymbol{r})|^2 d \boldsymbol{r}}{\int_{\Omega} u^2(\boldsymbol{r}) d \boldsymbol{r}} = \frac{\langle u , -\nabla^2 u \rangle}{\langle u , u \rangle},
\end{equation}
subject to the functions vanishing on the surface of the sphere $\Omega$, and where the inner product is the $L^2$ inner product on $\Omega$.
The stationary points of the Rayleigh quotient are the eigenfunctions of the operator $-\nabla^2$ with the chosen boundary conditions, and the value of the Rayleigh quotient at each stationary point is the corresponding eigenvalue.

This construction provides the smoothest possible set of basis functions under a certain notion of smoothness (in this case, the averaged value of the Laplace operator on the basis function). The Rayleigh quotient also provides a natural way to truncate the basis set by setting a smoothness threshold.
Furthermore, this method can be applied to high-order frameworks that define body-ordered functions in the $\Omega^{\otimes \nu}$ space, such as ACE~\cite{ace}, NICE~\cite{nice}, and MTP~\cite{mtp}, as high-order functions obeying the high-order equivalent of~\eqref{eq:rayleigh} can conveniently be expressed as products of the same first-order basis functions that are obtained by solving the first-order ($\nu=1$) equation on the 3D sphere~\cite{bigi2022smooth}.

Even more importantly, this choice of basis can be related to the regularization of the resulting properties~\cite{oord2020regularised}. We will exemplify this for an interatomic potential in the ACE framework. In ACE, an interatomic potential $V(\{ \boldsymbol{r}_i \}_{i=1}^N)$ is given by
\begin{equation}\label{eq:decomposition}
    V(\{ \boldsymbol{r}_i \}_{i=1}^N) = \sum_{i=1}^N V_i(\{ \boldsymbol{r}_j \}_{j=1}^{N_i}) = \sum_{i=1}^N \sum_{\nu=0}^{\nu_\mathrm{max}} V_i^{(\nu)} (\{ \boldsymbol{r}_j \}_{j=1}^{N_i}),
\end{equation}
where the term $V_i^{(\nu)}(\{ \boldsymbol{r}_j \}_{j=1}^{N_i})$ denotes the contribution to the atomic energy of atom $i$ at correlation order $\nu$, corresponding to body order $\nu+1$. 
The first equality results in an atom-wise decomposition of the potential, while the second equality expresses the decomposition of individual atomic energies into body-ordered components.
Finally, $V_i^{(\nu)} (\{ \boldsymbol{r}_j \}_{j=1}^{N_i})$ is parametrized as
\begin{equation}\label{eq:ace-single-bo}
    V_i^{(\nu)} (\{ \boldsymbol{r}_j \}_{j=1}^{N_i}) = \sum_b c_b B_b^{i, \nu, 00, +1} (\{ \boldsymbol{r}_\iota \}_{\iota=1}^\nu),
\end{equation}
where we have used the $B_b^{i, \nu LM \sigma}$ notation to denote an ACE B-basis function~\cite{ace} with index $b$, rotational symmetry equivalent to that of a spherical harmonic of degree $L$ and order $M$, and inversion symmetry $\sigma = \pm 1$. The $c_b$ coefficients are the fittable parameters of the model.

In this case, it is possible to equate the expectation value of the operator $\nabla^2$ to a regularization term.
Indeed, upon defining the $L^2$ inner product on $\Omega^\nu$ and considering the linear operator $\hat{L}$ associated with the corresponding Rayleigh quotient, the basis functions $B_b^{i, \nu, 00, +1}$ can be chosen as an orthonormal eigenbasis of $\hat{L}$, so that $\hat{L} B_b^{i, \nu, 00, +1} = E_b B_b^{i, \nu, 00, +1}$, where $E_b$ is the corresponding eigenvalue, equal to the Rayleigh quotient evaluated on that basis function. Therefore, we can write
\begin{equation}
    \begin{gathered}
        \langle V_i^{(\nu)}, \hat{L} V_i^{(\nu)} \rangle = \bigg\langle \sum_b c_b B_b^{i, \nu, 00, +1}, \hat{L} \sum_{b'} c_{b'} B_{b'}^{i, \nu, 00, +1} \bigg\rangle \\
        =\sum_b \sum_{b'} c_b c_{b'} E_{b'} \langle B_b^{i, \nu, 00, +1}, B_{b'}^{i, \nu, 00, +1} \rangle = \sum_b c_b^2 E_b,
    \end{gathered}
\end{equation}
obtained by using the eigenvalue equation and the orthonormality of the basis functions.
The last expression is exactly that of a regularization term in the loss of a linear fit with a diagonal regularization matrix with entries corresponding to $E_b$. Therefore, regularizing the potential is equivalent to imposing a penalty on high expectation values of the $\nabla^2$ operator, i.e. high waviness or low smoothness. It should be noted that the above construction is valid for any choice of linear operator in place of $\nabla^2$~\cite{oord2020regularised}.

\subsection{A physically inspired basis}

Although the construction in Ref.~\cite{bigi2022smooth} is elegant and allows to choose smooth functions, it results in the same level of smoothness being imposed within the whole atom-centered sphere, which is not a faithful physical prior.
Physical decay laws have been used widely in the modeling of long-range interactions, where sparsifying is essential due to the large amount of physical space to be covered efficiently. This section will apply the same principle to short-range descriptions of atomic environments. We argue that the corresponding law for short-range interactions is a negative exponential of the distance between atoms, reflecting the exponential decay of atomic orbitals.

In order to penalize non-smooth functions exponentially more as the distance between atoms increases, it is sufficient to multiply the numerator of the Rayleigh quotient by an exponential of the distance $r$ as follows:
\begin{equation}
\label{eq:3d_rayleigh}
    R = \frac{\int_{\Omega} |\nabla u(\boldsymbol{r})|^2 e^{r/r_0} d \boldsymbol{r}}{\int_{\Omega} u^2(\boldsymbol{r}) d \boldsymbol{r}}
\end{equation}
where $r_0$ is a lengthscale parameter.
Finding the functions that minimize this quotient requires solving an eigenvalue equation of the type $\hat{O}(u(\boldsymbol{r})) = E u(\boldsymbol{r})$, where $\hat{O}$ is a linear operator whose form is derived, together with the numerical solution of the resulting radial equation, in App.~\ref{app:physical-basis}.

\begin{figure*}[t]
  \centering
  \begin{subfigure}[b]{0.45\textwidth}
    \centering
    \includegraphics[width=\textwidth]{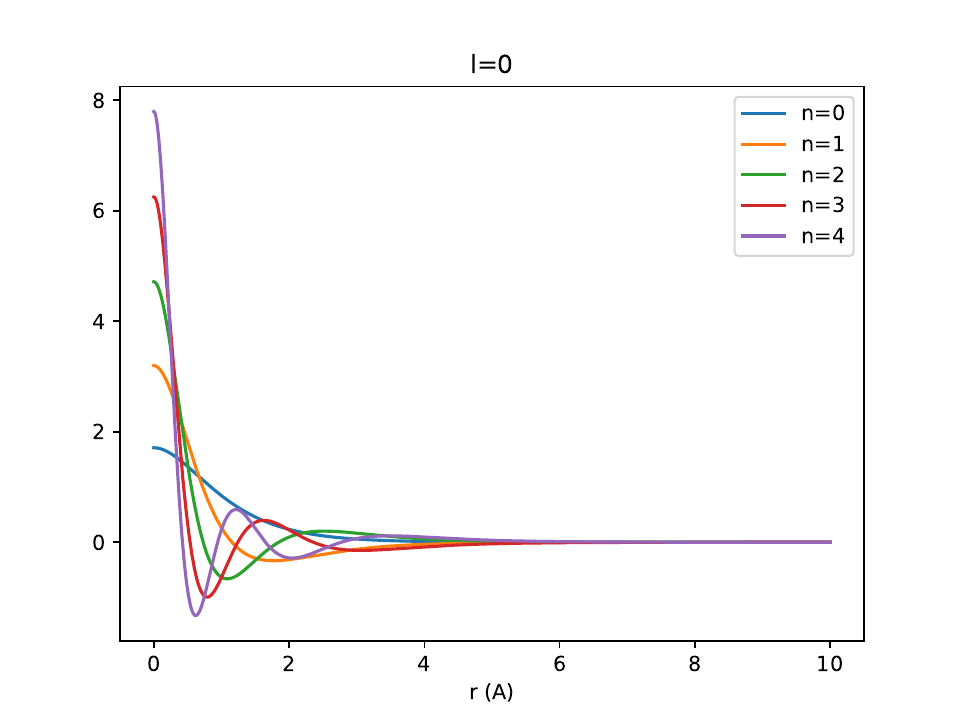}
  \end{subfigure}
  \begin{subfigure}[b]{0.45\textwidth}
    \centering
    \includegraphics[width=\textwidth]{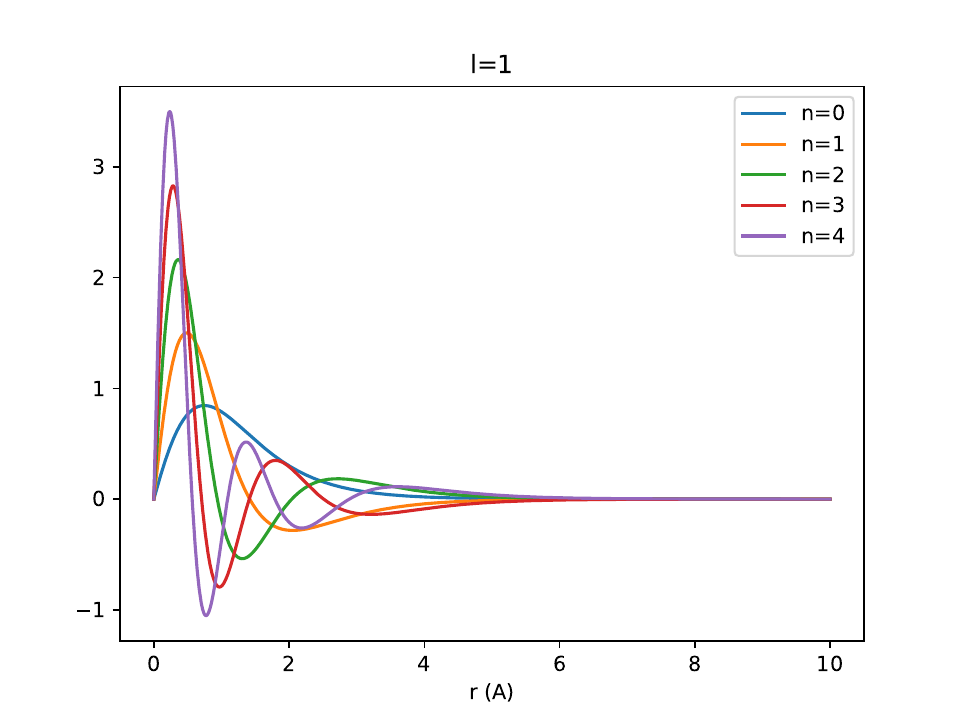}
  \end{subfigure}
  \vskip\baselineskip
  \begin{subfigure}[b]{0.45\textwidth}
    \centering
    \includegraphics[width=\textwidth]{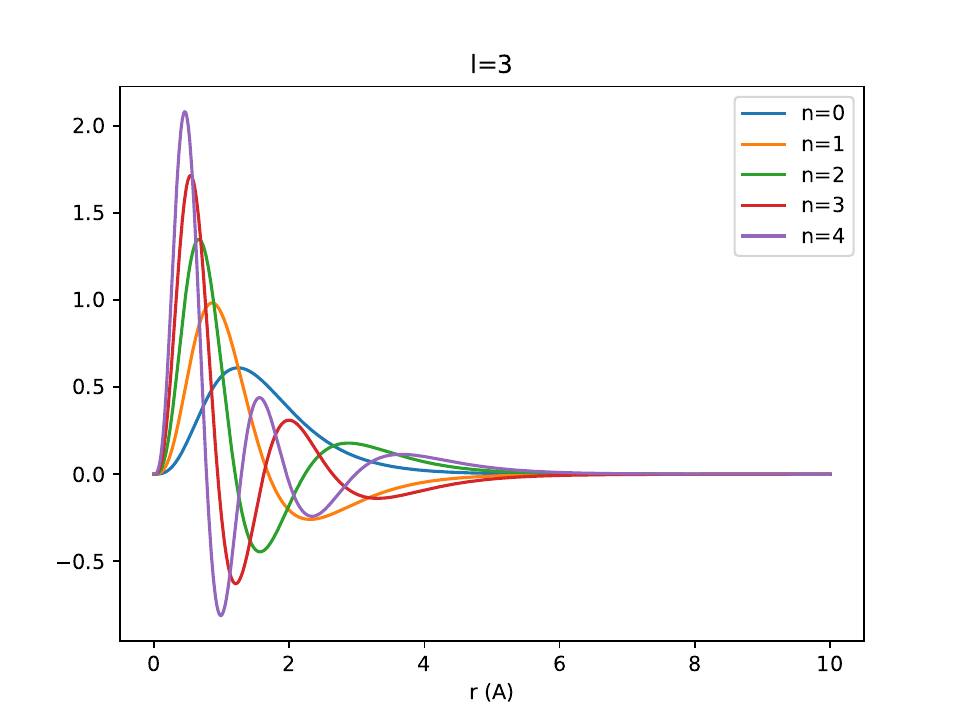}
  \end{subfigure}
  \begin{subfigure}[b]{0.45\textwidth}
    \centering
    \includegraphics[width=\textwidth]{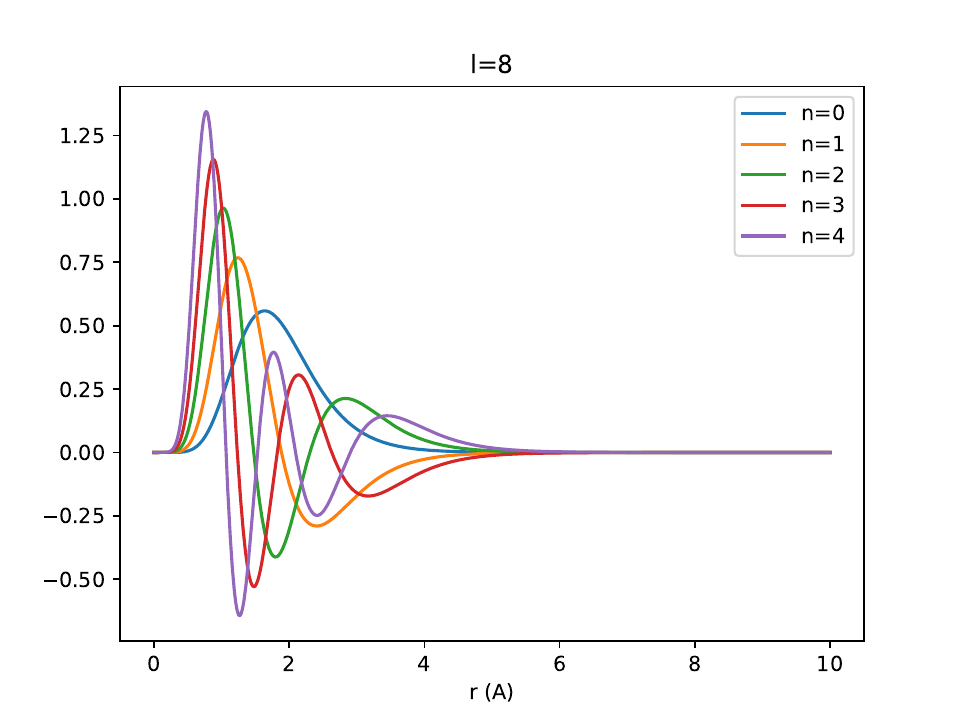}
  \end{subfigure}
  \caption{Radial basis functions $R_{nl}(r)$ of the physics-inspired basis for selected values of $l$ and $n=0,1,2,3,4$.}
  \label{fig:physical}
\end{figure*}

The solutions for the 3D case are products of a spherical harmonics function $Y_l^m$ and a radial basis function $R_{nl}$, i.e.,
\begin{equation}
u_{nlm} = R_{nl}(\boldsymbol{r}) Y_l^m(\boldsymbol{\hat{r}}).
\end{equation}
The radial basis functions for selected values of $l$ and $n$ are shown in Fig.~\ref{fig:physical}.
Exactly like the construction of Ref.~\cite{bigi2022smooth}, this physically inspired basis readily generalizes to the high-body-order descriptors that were considered in Sec.~\ref{sec:smoothness-and-regularization}. As such, it can be seen as a way to regularize body-ordered interactions of any order, provided that the values of the Rayleigh quotient (or, equivalently, the eigenvalues of $\hat{O}$) are used as the entries of the diagonal regularization matrix in a linear fit. 

\subsection{Efficient tensor products}

The tensor products we use in this work were introduced in Ref.~\cite{maennel2024complete}, and this section is intended as a brief primer on it. In order to introduce the tensor products, it will be useful to first introduce the concept of a ``compact equivariant representation''.

\paragraph{Compact equivariant representation}

It is known that if $\boldsymbol{x}^{(l)}$ is an SO(3)-equivariant object of degree $l$, then, for any $\boldsymbol{a}^{(l)}$ and $\boldsymbol{b}^{(l)}$, the tensor product $\boldsymbol{a}^{(l)} \otimes \boldsymbol{b}^{(l)}$ can be expressed in terms of irreducible representations as
\begin{equation}
     \bigoplus_{\lambda=0}^{L} \boldsymbol{c}^{(\lambda)},
\end{equation}
with $L=2l$.
Using the same idea in reverse~\cite{maennel2024complete}, a set of features from $\lambda=0$ to $\lambda=L$, where $L$ is even, can be expressed as a $(L+1) \times (L+1)$ matrix, which is simply the external product $\boldsymbol{a}^{(L/2)} \otimes \,  \boldsymbol{b}^{(L/2)}$, for some appropriate $\boldsymbol{a}^{(L/2)}$ and $\boldsymbol{b}^{(L/2)}$. For example, for $L=2$, we obtain the following layout:

\begin{center}
\begin{tikzpicture}[
  x=3.2mm, y=3.2mm,
  square/.style={draw, minimum size=3.2mm, inner sep=0pt, outer sep=0pt},
  lab/.style={font=\scriptsize, anchor=west}
]

\def\yzero{1.2}   %
\def\yone{0.0}    %
\def\ytwo{-1.2}   %

\node[square] at (2,\yzero) {};
\node[lab]    at (-4.0,\yzero) {$\lambda=0$};

\foreach \k in {0,1,2} { \node[square] at (1+\k,\yone) {}; }
\node[lab] at (-4.0,\yone) {$\lambda=1$};

\foreach \k in {0,1,2,3,4} { \node[square] at (\k,\ytwo) {}; }
\node[lab] at (-4.0,\ytwo) {$\lambda=2$};

\draw[->] (6,\yone) -- (8,\yone);

\begin{scope}[shift={(11.0,\yone)}]
  \foreach \r in {-1,0,1} {
    \foreach \c in {-1,0,1} {
      \node[square] at (\c,\r) {};
    }
  }
\end{scope}

\end{tikzpicture}
\end{center}

We define a \textit{compact} representation as the object on the right.

\paragraph{Tensor products in the compact representation}

It can be shown that, in the compact representation, where sets of irreducible representations are represented by square matrices, tensor products amount to a simple matrix multiplication between any two such square matrices~\cite{maennel2024complete}. This tensor product is entirely equivalent to taking the full tensor product of the two sets of irreducible spherical representations and truncating the result at a maximum degree of $\lambda=L$~\cite{maennel2024complete} (this truncation is already present almost universally in current equivariant architectures~\cite{duval2023hitchhiker}). Besides scaling as $\mathcal{O}(L^3)$ -- as opposed to $\mathcal{O}(L^6)$ for naive tensor products in the spherical representation -- this type of tensor product reduces to a single matrix multiplication and is therefore very suitable for modern machine learning frameworks and GPU execution.

We note in passing that in the compact representation it is possible to evaluate arbitrary equivariant functions, e.g. using their Taylor expansion, or any algorithm that can be expressed in terms of matrix products (cf. the scaling-and-squaring~\cite{mole-vanl03siamr}, as used to define exponential equivariant kernels in Ref.~\citenum{bigi+24jcp}).
We experimented with an architecture using an equivariant hyperbolic tangent nonlinear layer, but found the implementation to be plagued by numerical instabilities, and that the resulting form was less flexible than a sequence of compact equivariant products, that can effectively learn an arbitrary function rather than a specific nonlinear activation.

\begin{figure*}[t]
\includegraphics[width=\linewidth]{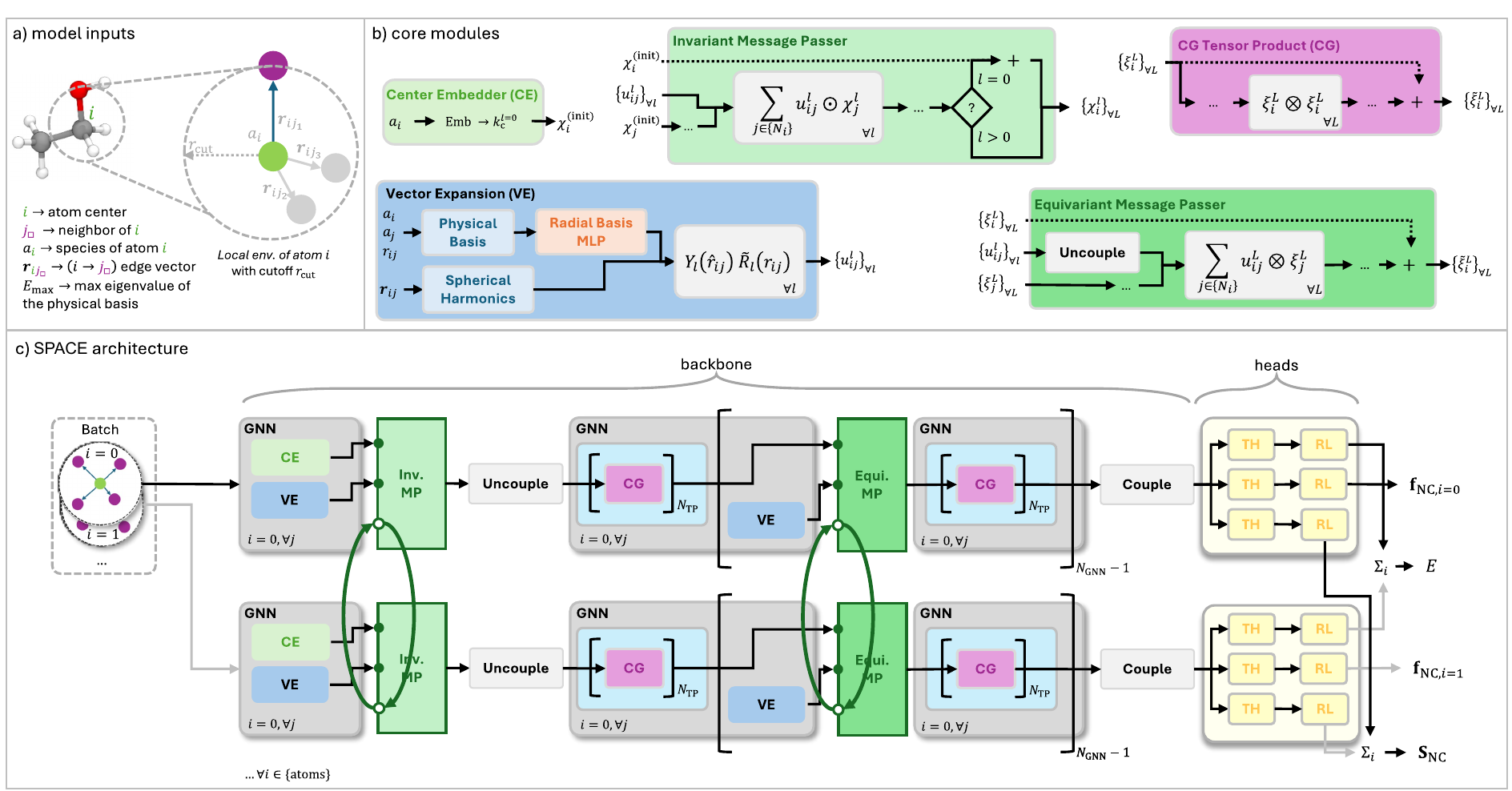}
\caption{{\small The SPACE architecture. a) The inputs to the model are atomistic systems of atomic positions and types (and cell parameters if periodic). Each system is decomposed into a set of local neighborhoods centered on each atom and defined within a cutoff. In SPACE, the single hyperparameter $E_{\text{max}}$ parametrizes the radial basis that expands the interatomic displacement vectors $\textbf{r}_{ij}$ and defines the maximum angular order and number of radial channels of the basis. b) The core modules of SPACE. Initial node features $\chi_i^{\text{(init)}}$ are given by an embedding of the central atom type (`Center Embedder'). The edge vectors in each atomic environment are expanded on a radial basis (`Physical Basis'), projected into a higher dimensional latent space by a non-linear multi-layer perceptron (`Radial Basis MLP'), and multiplied by the spherical harmonics to form a learnable spherical expansion $u_{ij}^l$. The `Invariant Message Passer' module combines the spherical expansion with initial neighbor features (neighbor type embeddings, $\chi_j^{\text{(init)}}$), aggregates over neighbors and updates the initial node features (center type embeddings) by a summation with invariant features (skip connection). The `Equivariant Message Passer' module takes a tensor product of a spherical expansion and neighbor features, both in the uncoupled basis, aggregates over neighbors, and updates node features for all group angular order $L$ by a direct equivariant sum. The `Clebsch-Gordan Tensor Product' module takes a set of node features in the compact (uncoupled) representation and takes the self-product: a matrix-matrix multiplication that doubles the body order of the features without the need for CG coefficients. c) The full SPACE architecture, from inputs to predictions. The local atomic environments from a batch of systems are processed in parallel by sequential GNN and message passing steps that form the backbone architecture. General GNN layers feature iterative application of CG tensor product modules (usually 6 per GNN) to node features, while message passing steps aggregate information from neighboring atoms, allowing the model to increase its receptive field beyond the cutoff. Backbone features in the uncoupled (compact) basis are coupled to form SO(3)-equivariant features, which are mapped by linear readout layers to the dimension of the corresponding target irreps to form the predictions. The figure shows an example of a SPACE-MLIP, where local features are mapped to form direct force ($l=1$) predictions, or aggreagted over center atoms to form total energy ($l=0$) and direct stress ($l=[0,2]$) predictions.} }
    \label{fig:space-arch}
\end{figure*}

\subsection{Overall architecture}

The architecture we use is inspired by the many architectures that use equivariant tensor products to build high-correlation-order representations within a graph neural network~\cite{mlace,mace,grace,ictp,cace,tace}. Compared to these architectures, we make a few concrete changes. (1) Our representation size varies with the order $\lambda$, and the representation size is dictated by the size of the truncated physically inspired basis for that $\lambda$. (2) All tensor products are performed in the compact representation. This implies transforming features to the compact representation (once for each message passing layer) and back from the compact representation (once at the end of the architecture). (3) In practice, we use a large number of tensor products (six or more per GNN layer, as opposed to the common choice of two~\cite{mace,grace}). This builds on the computational efficiency of tensor products in the compact representation, which are cheaper to evaluate than linear layers.

While (1) aims at improving the physical preconditioning and computational efficiency of the neural network, (2) and (3) achieve a complete equivariant representation up to very high correlation orders. Indeed, six consecutive tensor products reach a correlation order of $2^6 = 64$, and formal completeness up to this correlation order is guaranteed by the presence of skip connections and a full tensor product. While this is already larger than the typical neighbor count in an atomic environment, the correlation order reaches values of at least $2^{12} = 4096$ over multiple message-passing layers, which is much larger than the number of neighbors, even when considering the extended environment given by message passing (it suffices to consider that the correlation order scales exponentially with the number of message-passing layers, while the number of neighbors only cubically). This effectively creates a complete representation for any function that might be relevant for applications in atomic-scale modeling, without assuming convergence of the body-ordered expansion.

Finally, the architecture we propose is SO(3)-equivariant as opposed to O(3)-equivariant. The reason for relaxing inversion symmetry is that, in their naive form, O(3)-equivariant networks must contain separate even and odd representations, making their tensor products 4 times more expensive than SO(3) tensor products. Therefore, in practice, O(3)-equivariant networks instead often simply drop all odd representations~\cite{nequip,mace,allegro}, potentially at the expense of completeness and representation power. We instead choose not to enforce inversion symmetry (as done in, e.g., Refs.~\cite{frank2022so3krates} and~\cite{esen}), and we apply inversion augmentation at training time. We note that this is entirely harmless, as inversion symmetry can be recovered exactly at prediction time at the cost of a single additional evaluation.
An illustration of the architecture is given in Fig.~\ref{fig:space-arch}, and a more detailed description is given in App.~\ref{app:space-arch}.

\section{Conclusions}

The SPACE architecture has formally and practically a more benign scaling with the maximum angular momentum channel retained in the expansion, and allows the model to be scaled to include arbitrarily high orders of correlations, making it possible to express any function of the coordinates of all neighbors within the receptive field of the network. 
Expressive power and physical preconditioning of the radial basis however do not automatically translate into performance in regression tasks.
Careful testing across different types and sizes of datasets is needed to assess the concrete impact of these architectural choices.
For the moment, we present SPACE as a demonstration that the design space of equivariant architectures still contains poorly explored regions, that may provide a better balance between expressivity and computational effort. 

\section{Data availability}

An implementation of SPACE is openly available as part of the \texttt{metatrain} package~\cite{bigi+26jcp}.

\appendix
\onecolumngrid
\section{Construction of the physical basis}\label{app:physical-basis}

Integrating the numerator of the Rayleigh quotient in Eq.~\eqref{eq:3d_rayleigh} by parts, and using the fact that $u(\boldsymbol{r})$ vanishes on the boundary of $\Omega$, leads to
\begin{equation}
    R = \frac{\int_{\Omega} u(\boldsymbol{r}) \left(-\nabla^2 u(\boldsymbol{r}) - \frac{1}{r_0} \frac{\partial u}{\partial r}\right) e^{r/r_0} d \boldsymbol{r}}{\int_{\Omega} u^2(\boldsymbol{r}) d \boldsymbol{r}}.
\end{equation}
The integrand of the numerator has the form $u(\boldsymbol{r}) \hat{O}(u(\boldsymbol{r}))$, where the linear operator $\hat{O}$ can be identified as
\begin{equation}
    \hat{O} = e^{r/r_0}\left(-\nabla^2 - \frac{1}{r_0} \frac{\partial}{\partial r}\right).
\end{equation}
The stationary points of the Rayleigh quotient are therefore the solutions of the eigenvalue equation $\hat{O} u(\boldsymbol{r}) = E u(\boldsymbol{r})$, or
\begin{equation}
    e^{r/r_0}\left(-\nabla^2 u(\boldsymbol{r}) - \frac{1}{r_0} \frac{\partial u(\boldsymbol{r})}{\partial r}\right) = E u(\boldsymbol{r}),
\end{equation}
where the eigenvalue $E$ equals the value of the Rayleigh quotient for the corresponding eigenfunction.

Since the angular dependence of $\hat{O}$ is entirely contained in the Laplace operator, the solutions factorize as $u(\boldsymbol{r}) = R(r) Y_l^m(\theta, \phi)$, where $R(r)$ must satisfy
\begin{equation}
    e^{r/r_0} \left( -\frac{d^2 R}{dr^2} - \frac{2}{r} \frac{dR}{dr} + \frac{l(l+1)}{r^2} R - \frac{1}{r_0} \frac{dR}{dr} \right) = ER.
\end{equation}
Substituting $x = r/r_0$ results in
\begin{equation}
    e^x \left( -\frac{d^2 R}{dx^2} - \frac{2}{x} \frac{dR}{dx} + \frac{l(l+1)}{x^2} R - \frac{dR}{dx} \right) = r_0^2 E R.
\end{equation}
The solutions of this equation are found numerically, by expanding $R(x)$ on a set of trigonometric functions within the interval $[0, 10]$, as the solutions are observed to vanish almost completely well before $x=10$. The trigonometric functions are chosen to vanish at $x=10$ and, for $l>0$, also at $x=0$; for $l=0$, whose radial functions do not vanish at the origin, they are instead non-zero at $x=0$. Since the solutions are independent of $r_0$ (only the eigenvalues scale by a factor of $r_0^2$), they can be conveniently used to generate radial basis functions for any value of $r_0$. The truncation of the resulting basis is discussed in App.~\ref{app:space-arch}.

The Python and C++ programs used to solve the eigenvalue equations can be found at \url{https://github.com/frostedoyster/physical_basis}.

\section{SPACE architecture details}~\label{app:space-arch}

This section covers in more detail the SPACE architecture summarized in the main text. The full architecture is shown in Figure~\ref{fig:space-full}.

\subsection{The physical basis in practice}

The size of the radial basis that expands the edge vectors in each atomic environment is controlled by a single hyperparameter, $E_{\text{max}}$, corresponding to the maximum eigenvalue of the physical basis. Increasing $E_{\text{max}}$ in general increases the maximum angular order of the basis and the number of radial channels for each $l$.

The value of the hyperparameter $E_{\text{max}}$ should be chosen to cover the maximum angular order of the physical property being targeted, for example, with guidance from Table~\ref{tab:choosing_emax}, which shows the maximum angular order $l_{\text{max}}$ and size of the radial basis $n_{\text{max}}^l$ obtained given a choice of $E_{\text{max}}$. For instance, when learning dipoles (a vectorial target with maximum angular order $l=1$), one should set at least $E_{\text{max}} \ge 12.97$, though increasing the maximum eigenvalue beyond this will give a larger and more expressive basis. 

\begin{table}
\begin{tabular}{cc c l}
\toprule
\multicolumn{2}{c}{$E_{\text{max}} \in [\text{low}, \text{high})$} & \multicolumn{1}{c}{$l_{\text{max}}$} & \multicolumn{1}{c}{$\{n_{\text{max}}^{l} \}$} \\
low & high & \\
\midrule
    $8.00$ & $12.97$ & $0$ & $[1]$ \\
    $12.97$ & $16.93$ & $1$ & $[1, 1]$ \\
    $16.93$ & $22.36$ & $1$ & $[2, 1]$ \\
    $22.36$ & $25.64$ & $2$ & $[2, 1, 1]$ \\
    $25.64$ & $31.53$ & $2$ & $[2, 2, 1]$ \\
    $31.53$ & $35.32$ & $2$ & $[3, 2, 1]$ \\
    $35.32$ & $38.13$ & $3$ & $[3, 2, 1, 1]$ \\
    $38.13$ & $43.03$ & $3$ & $[3, 2, 2, 1]$ \\
    $43.03$ & $50.96$ & $3$ & $[3, 3, 2, 1]$ \\
    $50.96$ & $51.93$ & $3$ & $[4, 3, 2, 1]$ \\
    $51.93$ & $54.37$ & $4$ & $[4, 3, 2, 1, 1]$ \\
    $54.37$ & $58.57$ & $4$ & $[4, 3, 2, 2, 1]$ \\
    $58.57$ & $65.21$ & $4$ & $[4, 3, 3, 2, 1]$ \\
    $65.21$ & $72.21$ & $4$ & $[4, 4, 3, 2, 1]$ \\
    $72.21$ & $74.37$ & $5$ & $[4, 4, 3, 2, 1, 1]$ \\
    $74.37$ & $75.25$ & $5$ & $[4, 4, 3, 2, 2, 1]$ \\
    $75.25$ & $78.03$ & $5$ & $[5, 4, 3, 2, 2, 1]$ \\
    $78.03$ & $83.74$ & $5$ & $[5, 4, 3, 3, 2, 1]$ \\
    $83.74$ & $92.23$ & $5$ & $[5, 4, 4, 3, 2, 1]$ \\
    $92.23$ & $96.20$ & $5$ & $[5, 5, 4, 3, 2, 1]$ \\
    $96.20$ & $98.12$ & $6$ & $[5, 5, 4, 3, 2, 1, 1]$ \\
    $98.12$ & $101.38$ & $6$ & $[5, 5, 4, 3, 2, 2, 1]$ \\
    $101.38$ & $104.43$ & $6$ & $[5, 5, 4, 3, 3, 2, 1]$ \\
    $104.43$ & $106.38$ & $6$ & $[6, 5, 4, 3, 3, 2, 1]$ \\
    $106.38$ & $113.71$ & $6$ & $[6, 5, 4, 4, 3, 2, 1]$ \\
    $113.71$ & $123.87$ & $6$ & $[6, 5, 5, 4, 3, 2, 1]$ \\
    $123.87$ & $124.10$ & $7$ & $[6, 5, 5, 4, 3, 2, 1, 1]$ \\
    $124.10$ & $125.62$ & $7$ & $[6, 6, 5, 4, 3, 2, 1, 1]$ \\
    $125.62$ & $128.57$ & $7$ & $[6, 6, 5, 4, 3, 2, 2, 1]$ \\
    $128.57$ & $133.03$ & $7$ & $[6, 6, 5, 4, 3, 3, 2, 1]$ \\
    $133.03$ & $138.50$ & $7$ & $[6, 6, 5, 4, 4, 3, 2, 1]$ \\
    $138.50$ & $139.48$ & $7$ & $[7, 6, 5, 4, 4, 3, 2, 1]$ \\
    $139.48$ & $148.52$ & $7$ & $[7, 6, 5, 5, 4, 3, 2, 1]$ \\
    $148.52$ & $155.24$ & $7$ & $[7, 6, 6, 5, 4, 3, 2, 1]$ \\
    $155.24$ & $156.87$ & $8$ & $[7, 6, 6, 5, 4, 3, 2, 1, 1]$ \\
    $156.87$ & $159.57$ & $8$ & $[7, 6, 6, 5, 4, 3, 2, 2, 1]$ \\
    $159.57$ & $160.87$ & $8$ & $[7, 6, 6, 5, 4, 3, 3, 2, 1]$ \\
    $160.87$ & $163.62$ & $8$ & $[7, 7, 6, 5, 4, 3, 3, 2, 1]$ \\
    $163.62$ & $169.39$ & $8$ & $[7, 7, 6, 5, 4, 4, 3, 2, 1]$ \\
    $169.39$ & $177.37$ & $8$ & $[7, 7, 6, 5, 5, 4, 3, 2, 1]$ \\
    $177.37$ & $188.16$ & $8$ & $[8, 7, 6, 6, 5, 4, 3, 2, 1]$ \\
    $188.16$ & $190.32$ & $8$ & $[8, 7, 7, 6, 5, 4, 3, 2, 1]$ \\
    $190.32$ & $191.83$ & $9$ & $[8, 7, 7, 6, 5, 4, 3, 2, 1, 1]$ \\
    $191.83$ & $194.34$ & $9$ & $[8, 7, 7, 6, 5, 4, 3, 2, 2, 1]$ \\
    $194.34$ & $198.08$ & $9$ & $[8, 7, 7, 6, 5, 4, 3, 3, 2, 1]$ \\
    $198.08$ & $202.53$ & $9$ & $[8, 7, 7, 6, 5, 4, 4, 3, 2, 1]$ \\
    $202.53$ & $203.34$ & $9$ & $[8, 8, 7, 6, 5, 4, 4, 3, 2, 1]$ \\
    $203.34$ & $210.49$ & $9$ & $[8, 8, 7, 6, 5, 5, 4, 3, 2, 1]$ \\
    $210.49$ & $220.06$ & $9$ & $[8, 8, 7, 6, 6, 5, 4, 3, 2, 1]$ \\
    $220.06$ & $221.40$ & $9$ & $[8, 8, 7, 7, 6, 5, 4, 3, 2, 1]$ \\
    $221.40$ & $229.09$ & $9$ & $[9, 8, 7, 7, 6, 5, 4, 3, 2, 1]$ \\
    $229.09$ & $230.51$ & $10$ & $[9, 8, 7, 7, 6, 5, 4, 3, 2, 1, 1]$ \\
    $230.51$ & $232.68$ & $10$ & $[9, 8, 7, 7, 6, 5, 4, 3, 2, 2, 1]$ \\
    $232.68$ & $236.37$ & $10$ & $[9, 8, 8, 7, 6, 5, 4, 3, 3, 2, 1]$ \\
    $236.37$ & $241.23$ & $10$ & $[9, 8, 8, 7, 6, 5, 4, 4, 3, 2, 1]$ \\
    $241.23$ & $247.75$ & $10$ & $[9, 8, 8, 7, 6, 5, 5, 4, 3, 2, 1]$ \\
    $247.75$ & $249.10$ & $10$ & $[9, 8, 8, 7, 6, 6, 5, 4, 3, 2, 1]$ \\
    $249.10$ & $256.37$ & $10$ & $[9, 9, 8, 7, 6, 6, 5, 4, 3, 2, 1]$ \\
    $256.37$ & $267.59$ & $10$ & $[9, 9, 8, 7, 7, 6, 5, 4, 3, 2, 1]$ \\
    $267.59$ & $270.23$ & $10$ & $[9, 9, 8, 8, 7, 6, 5, 4, 3, 2, 1]$ \\
    $270.23$ & $271.55$ & $10$ & $[10, 9, 8, 8, 7, 6, 5, 4, 3, 2, 1]$ \\
\bottomrule
\end{tabular}
\caption{The maximum angular order, $l_{\text{max}}$, and number of radial channels per angular order, $\{n_{\text{max}}^{l} \}$, tabulated as a function of the maximum eigenvalue $E_{\text{max}}$ of the physical basis. Ranges of $E_{\text{max}}$ are given up to highest value that corresponds to $l_{\text{max}} = 10$.}
\label{tab:choosing_emax}
\end{table}

The reason the physical basis constructed initially must cover the full range of angular orders present in the targets is due to the truncation of angular features implied in the CG tensor product in the compact basis: features of angular order up to $L$ do not increase beyond this order by construction.

For a given central atom $i$, the radial basis for each edge vector $\textbf{r}_{ij}$ between it and each neighbor $j$ is constructed by the `Physical Basis' module (Figure~\ref{fig:space-full}b), using a distance that is scaled according to the covalent radii of the corresponding atom types $a_i$ and $a_j$. A multi-layer perceptron (MLP) projects the radial basis of dimension $n_{\text{max}}^l$ into the latent space of dimension $k_{\text{c}}^l = d_{\text{elem}} \times n_{\text{max}}^l$, where $d_{\text{elem}}$ is a fixed size embedding factor. A vector expansion of the embedded radial basis with the spherical harmonics produces a learned physical basis $u_{ij}^l$ that form the features that propagate through the network.

\subsection{Efficient tensor products}

An efficient implementation of the Clebsch-Gordan tensor product in the uncoupled/compact basis is what allows SPACE to reach effectively complete body ordered features~\cite{maennel2024complete}. The heart of SPACE architecture is the `Clebsch-Gordan Iterator' module (Figure~\ref{fig:space-full}c) that iteratively performs self-products of the inputs in the uncoupled basis. Each CG iteration multiplies the body order by two, such that in general the `CG Iterator' multiplies it by $2^{N_{\text{TP}}}$, where $N_{\text{TP}}$ is the number of tensor products performed per GNN layer.

\begin{figure}[H]
\includegraphics[width=\linewidth]{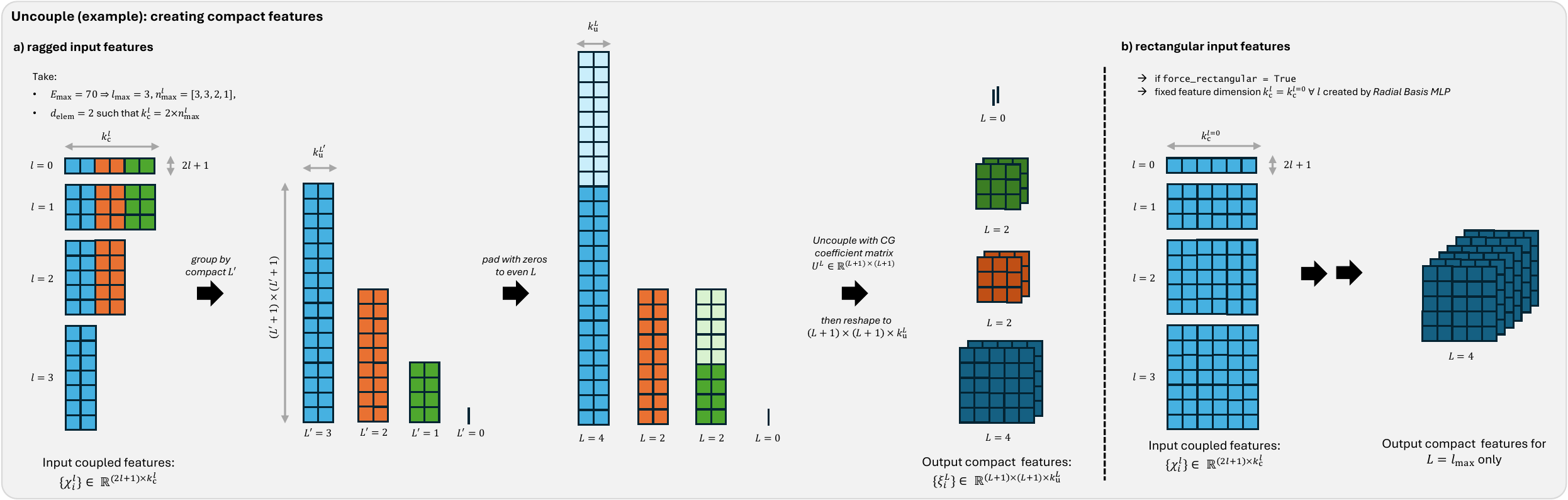}
\caption{
    Visual example of how a) \textit{ragged} and b) \textit{rectangular} features in the coupled basis are transformed into the compact representation in the uncoupled basis, ready for the efficient CG tensor product.
}
    \label{fig:compact}
\end{figure}

An example of how compact representations of features in the uncoupled basis are created from features in the coupled basis is shown in Figure~\ref{fig:compact}. 

In general, the radial basis returned by the `Physical Basis' module are ragged across different values of $l$, which propagates through to the embedded space when passed through `Radial Basis MLP' (Figure~\ref{fig:compact}a). 
Starting with $l=l_{\text{max}}$ and working down through the angular orders until $l=0$, features are collected into groups labelled by the maximum order they contain $L'$. In this example, the $l=0$ features are split into all groups, but because of the same size of the radial basis for $l=0$ and $l=1$, there exists no features in the grouped $L'=0$ features.
Grouped features are then padded with zeros in the angular components to give them the dimension of grouped features with even $L$: this is because compact representations in the uncoupled basis do not exist for odd $L$. Features are uncoupled using the corresponding CG coefficient matrix, resulting in compact features of size (for a single atom) $(L+1) \times (L+1) \times k_{\text{u}}^L$, where $k_{\text{u}}^L$ is the $L$-dependent feature size in the uncoupled basis, given by the difference between feature sizes consecutive in $l$ in the coupled basis: $k_{\text{u}}^L = k_{\text{c}}^L - k_{\text{c}}^{L+1}$.

Input coupled features can also be enforced to be not ragged, instead \textit{rectangular}, by action of the `Radial Basis MLP' module (Figure~\ref{fig:compact}b). In this case, a single large compact representation is created of maximum group angular order $L = l_{\text{max}}$.

In both cases, the dense features in the coupled representation enter the Clebsch-Gordan tensor product modules as is, and self-products are executed by simple matrix multiplication with no further involvement of CG coefficient matrices from within the GNN backbone (until the final coupling at the heads).

\subsection{Message passing modules}

The initial node features are given by an embedding of the central atom type by the `Center Embedder'.
In the `Invariant Message Passer' module (the first message passing step of the network), the atom type embeddings of atoms neighboring the central atom form the initial messages. A product of these messages with the embedded physical basis, an aggregation over neighbors, and a skip connection with the central atom embeddings updates the node features in the coupled basis (Figure~\ref{fig:space-full}d). These are then uncoupled before into the compact representation before passage through the CG Iterator.

The `Equivariant Message Passer' module, on the other hand, takes a tensor product between a vector expansion an the uncoupled node features of neighboring atoms $\{ \xi_j^L \}$ for each group angular order $L$, along with aggregation over neighbors. A direct sum by skip connection then updates the node features for the central atom (Figure~\ref{fig:space-full}d) for each group angular order $L$.

\subsection{Full backbone and prediction layers}

\begin{figure}[H]
\includegraphics[width=\linewidth]{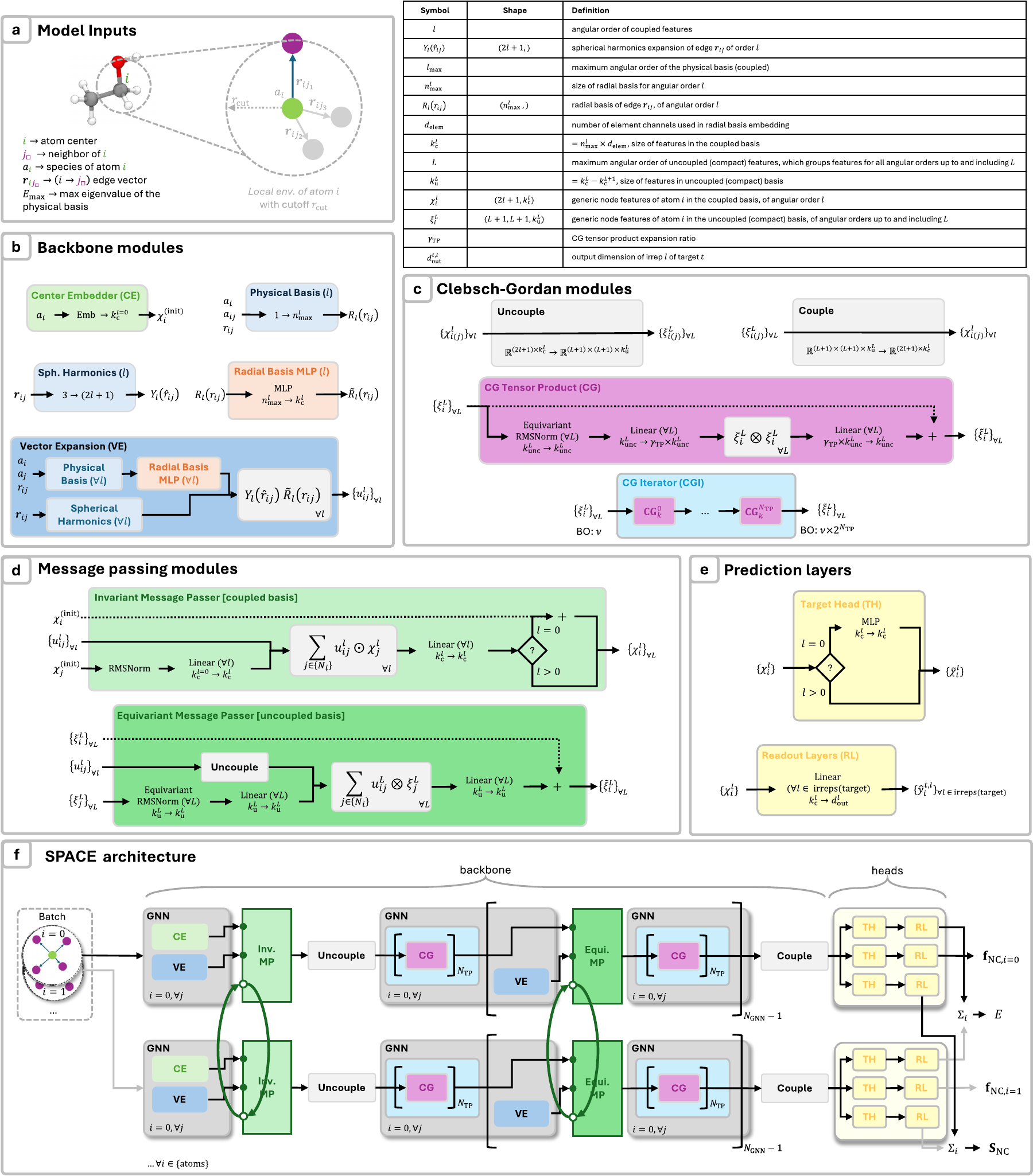}
\caption{The SPACE architecture}
    \label{fig:space-full}
\end{figure}

The message passing and CG iteration modules are combined in sequence to form the backbone of the SPACE architecture for a specified number of GNN layers. The final backbone features are transformed to the coupled basis, and those with angular orders that match the angular decomposition of the targets being learned are selected. 
For invariant (scalar, $l=0$) irreps of the target (i.e. the energy, or isotropic part of the stress), the target head is a non-linear MLP (Figure~\ref{fig:space-full}e). For covariant ($l>0$) target irreps, no head is applied. The last-layer features are then mapped from their respective feature dimension $k_{\text{c}}^l$ to the output dimension to form the predictions.

\end{document}